\documentclass[prl,twocolumn,bibnotes,showpacs,amsbsy,floatfix,amsbsy,floatfix,superscriptaddress]{revtex4-1}
\usepackage{epsfig,color}
\usepackage[cp1250]{inputenc}
\usepackage{amsmath}
\usepackage{amsfonts}
\usepackage{color}
\usepackage{latexsym}

\usepackage{natbib}

\begin{document}

\title{Long-Distance Entanglement of Soliton Spin Qubits in Gated Nanowires}

\author{Pawe\l{} Szumniak}
\affiliation{Department of Physics, University of Basel, Klingelbergstrasse 82, 4056 Basel, Switzerland}
\affiliation{AGH University of Science and Technology, Faculty of
Physics and Applied Computer Science,\\
al. Mickiewicza 30, 30-059 Krak\'ow, Poland}
\author{Jaros\l{}aw Paw\l{}owski}
\affiliation{AGH University of Science and Technology, Faculty of
Physics and Applied Computer Science,\\
al. Mickiewicza 30, 30-059 Krak\'ow, Poland}
\author{Stanis\l{}aw Bednarek}
\affiliation{AGH University of Science and Technology, Faculty of
Physics and Applied Computer Science,\\
al. Mickiewicza 30, 30-059 Krak\'ow, Poland}
\author{Daniel Loss}
\affiliation{Department of Physics, University of Basel, Klingelbergstrasse 82, 4056 Basel, Switzerland}
\date{\today}
\begin{abstract}
We investigate numerically charge, spin, and entanglement dynamics of two electrons confined in a gated semiconductor nanowire. The electrostatic coupling between electrons in the nanowire and the charges in the metal gates leads to a self-trapping of the electrons which results in soliton-like properties. 
We show that the interplay of an all-electrically controlled coherent transport of the electron solitons and of the exchange interaction can be used to realize ultrafast SWAP and entangling $\sqrt{\text{SWAP}}$ gates for distant spin qubits. We demonstrate that the latter gate can be used to generate a maximally entangled spin state of spatially separated electrons. The results are obtained by quantum mechanical time-dependent calculations with exact inclusion of electron-electron correlations.

\end{abstract}


\maketitle
{\it Introduction.} One of the most striking manifestations of the quantum laws of physics is  entanglement~\cite{PhysRev.47.777,RevModPhys.81.865}. 
The ability to entangle qubits is also an essential ingredient for  quantum computation~\cite{nielsen2000quantum}. The spins of electrons confined in an array of electrostatically defined quantum dots (QD) emerged as a promising candidate for encoding quantum bits of information~\cite{3, awschalom2013quantum}. Spin qubits weakly interacting with their environment can be electrically controlled and show potential in scalability~\cite{CKRevl}. Recent experiments have demonstrated fast and precise manipulation, initialization, and measurement of spin qubits confined in lateral QDs~\cite{9,6,s18,PhysRevLett.107.146801,s14,PhysRevLett.112.026801,shi2014fast,veldhorst2014addressable} and nanowire QDs~\cite{s1,*s3}.
Furthermore long spin decoherence times have been reported, reaching $\sim 200ns$~\cite{s1} for InAs QDs and $\sim 270\mu$s~\cite{ms} for GaAs QDs. 

However, the ability to couple and entangle  solid state spin qubits over long distances, which is essential for realizations of scalable quantum computer architectures and for applications of fault-tolerant quantum error correction (QEC) schemes, still seems to be one of the key challenges to overcome.
First encouraging steps toward coupling remote spin qubits via microwave cavities~\cite{PhysRevLett.83.4204, PhysRevB.74.041307,PhysRevB.77.045434,PhysRevLett.108.190506} have been recently reported~\cite{PhysRevLett.108.046807, petersson2012circuit, PhysRevLett.108.190506}. 
Coherent long-range spin qubit coupling based on cotunneling phenomena~\cite{PactStano} has been demonstrated for an array consisting of three 
quantum dots~\cite{braakman2013long, PhysRevLett.112.176803}.
There are also several interesting proposals for coupling spatially separated QD spin qubits, 
{\it e.g.} using ferromagnets~\cite{PhysRevX.3.041023}, floating gates~\cite{PhysRevX.2.011006}, Majorana edge modes~\cite{PhysRevLett.106.130505,PhysRevLett.107.210502,*PhysRevB.86.104511} or superconductors~\cite{PhysRevB.62.13569,*PhysRevLett.111.060501}.

Another promising platform useful for coupling spin qubits are mobile electrons shuttled by surface acoustic waves~\cite{SAW1,*SAW2,s19} or flying qubits~\cite{yamamoto2012electrical}, however, spin entanglement of such moving electrons has not been reported so far.

In this paper we propose an all-electrically controlled and ultrafast method for realization of the SWAP and $\sqrt{\text{SWAP}}$ gates and for generating maximally entangled spin states of spatially separated electrons. Our scheme does not require coupling with an additional external system - `quantum bus' -  which may simplify its implementation. The proposed scheme is based on the interplay between the exchange interaction and on-demand coherent transport of self-trapped electron solitons~\cite{yano1992single, PhysRevB.72.075319} confined in gated semiconductor nanostructures.
Exchange interaction, which has a short-range character, limits the ability to couple spatially separated spin qubits confined in stationary QDs.
However, for mobile electron solitons this is not the case.
Self-trapping allows for transporting spatially separated initially not entangled electrons to the region where they can entangle their spins due to the exchange interaction
and finally be separated and transported back to distant regions as an entangled entity. 
To be specific we present the results for structurally defined InAs  nanowires. However, one can expect qualitatively similar results for electrostatically defined quantum wires in 2DEG/2DHG systems as proposed e.g. in Refs.~\cite{PhysRevLett.100.126805}, and for different materials. 
One can integrate  SWAP and $\sqrt{\text{SWAP}}$ gates with all-electrically controlled single ultrafast quantum logic gates~\cite{14,HH1} which can be arranged in a 2D scalable register (Ref.~\cite{HH2}) and be selectively manipulated. Such an architecture may be particularly suitable for implementation of 
powerful QEC surface codes~\cite{PhysRevLett.98.190504}.


{\it Model.} We consider two electrons confined in an InAs semiconductor nanowire covered by  $7$ electrodes $e_{L}$, $e_{R}$, $e_{J}$, and $e_{1-4}$, to which the voltages $V_{L}$, $V_{R}$, $V_{J}$, and $V_{1-4}$, are applied~\footnote{Since the metal gate is in contact with an undoped semiconductor, the Schottky barrier $V_B$  should be taken into account with mV accuracy. Therefore, the potential applied to the gates is equal to $V_i \rightarrow V_i+V_B$. $V_B$ should be determined experimentally for a certain structure.}, respectively. The inter-electrode distance is about 10nm. The radius of the nanowire is $l=5$nm. The nanowire is separated from the metal by a dielectric material (InAlAs)~\footnote{We assume that value of the dielectric constant of the material surrounding the nanowire is equal or close to that of the nanowire.}, and the distance from the center of the nanowire to the metal is $d=15$nm.
\begin{figure}[ht!]
\centering
\includegraphics[width=8.6cm]{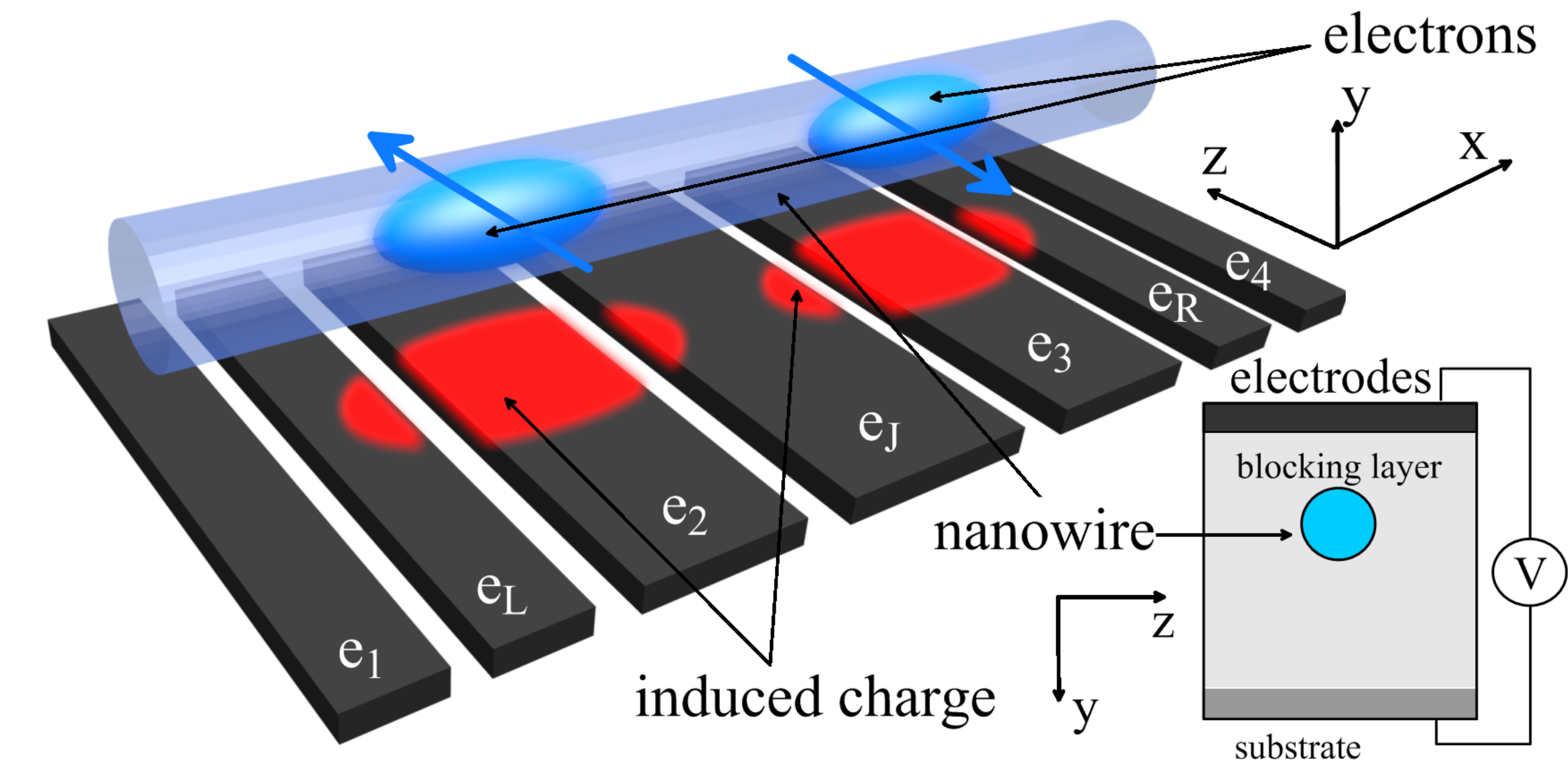}
\caption{\label{fig:nan} (color online) Schematic view of the nanostructure and its cross-section. The nanowire is covered by 7 electrodes which are labeled by $e_{1-4}$ and $e_{L, R, J}$. Two electrons with opposite spins (blue arrows) are confined in the nanowire. The  positive charge induced on the gate surface is marked by red areas.}
\end{figure}
The presented system can be described by the quasi-1D Hamiltonian, 
\begin{equation}
  {H}=\sum_{j=1,2}\left(\frac{{p}_{j}^2}{2m}+V_{conf}\big(x_j, \rho(x,t)\big)\right)+V_C^{1D}(x_1,x_2)+{H}_{BR},
\label{eq:Htot}
\end{equation}
where ${p}_{j}=-i\hbar{\partial}/{\partial x_j}$ 
is the x-component of the momentum operator for the $j$-th electron, $j=1, 2$, $m=0.023m_0$ is the effective mass of the electrons confined in the InAs nanowire, $m_0$ is the free electron mass. The two-electron wave function is represented as~\cite{27}
\begin{equation}
	\Psi(x_1,x_2,t)=\big(\psi_{\uparrow\uparrow},\psi_{\uparrow\downarrow},\psi_{\downarrow\uparrow},
\psi_{\downarrow\downarrow}\big)^T.
\label{eq:Psitot}
\end{equation}
It has to be antisymmetric with respect to simultaneous exchange of the space and spin coordinates: $\psi_{ij}(x_1,x_2,t)=-\psi_{ji}(x_2,x_1,t)$, where $i,j=\uparrow,\downarrow$ indicate the spin projections of the first and the second electron on the z axis.

The  electrons in the nanowire induce a positive charge on the surface of the metal electrodes which in turn leads to the self-confinement of the electron wave function along the wire. This wave function has soliton-like properties~\cite{PhysRevB.72.075319}: It can be
transported in the form of a stable wave packet which maintains its shape
while traveling. Moreover, it can reflect or pass through
obstacles (potential barriers or wells) with 100\% probability
while preserving its shape. This effect can be exploited to realize on-demand transport of self-confined electrons, whose motions can be fully controlled by geometry and voltages applied to the electrodes~\cite{PhysRevLett.100.126805}.
In the general case (e.g. arbitrary geometry of the electrodes) the induced self-confining potential is determined using the Poisson-Schr\"odinger self-consistent scheme which was described in detail in Refs.~\cite{PhysRevLett.100.126805,14,HH2}. However, since in the considered structure electrodes form almost uniform wide plates (the inter-electrode distance is about 10nm), in order to determine the induced potential, we can use instead the image charge technique ~\cite{PhysRevB.72.075319},
\begin{equation}
    V_{ind}(x, t)=\frac{e}{4\pi\varepsilon\varepsilon_0}\int dx'\frac{\rho(x',t)}{\sqrt{(x-x')^2+4d^2}}. 
\label{eq:Vind}
\end{equation}
This greatly simplifies and speeds up the numerical calculations.
Quantum calculations~\cite{PhysRevB.73.155318} indicate that it is a good approximation of the actual response potential of the electron gas. 
Here, $\varepsilon=14.3\varepsilon_0$ is the dielectric constant for the InAs nanowire \cite{winkler2003spin}.
The two-electron charge density is defined as $\rho(x,t)=e\int dx'\left(|\Psi(x,x',t)|^2+|\Psi(x',x,t)|^2\right)$,
where $|\Psi(x_1,x_2,t)|^2=\sum_{i,j=\uparrow,\downarrow}|\psi_{ij}(x_1,x_2,t)|^2$.
The voltages applied to the electrodes generate an additional electrostatic potential in the nanowire region, $\phi_0(x, y_0, z_0)$, which we determine by solving the Laplace equation, $\nabla^2\phi_0(x, y, z)=0$, under the  conditions $\phi_0(x, y_0, z)=V_i$, where $V_i$ is the voltage applied to the i-th electrode. Thus, according to the superposition principle, the total confinement potential can be expressed as $V_{conf}\big(x, \rho(x,t)\big)=-|e|V_{ind}(x, t)-|e|\phi_0(x,y_0,z_0)$. 

The electric field, $E_y$, generated by the electrodes and the induced charges are also the source of the Rashba spin-orbit interaction  in the nanowire, $H_{R}^j=\alpha E_yk_{x_j}\sigma_z$ \cite{BR1, *BR2}. However, for a chosen orientation of the electrodes, wire, and initial electron spin (either up or down along the z direction) the motion of the electrons along the wire does not induce spin rotations. Such spin-orbit interaction only slightly increases spin swap times.  
Furthermore we assume that the nanowire is grown along $[111]$ crystallographic direction which allows us to neglect the Dresselhaus spin-orbit(DSO) interaction \cite{winkler2003spin} which can affect slightly gate fidelity \cite{PhysRevB.82.165316}.

The effective 1D Coulomb interaction between charge carriers in  a nanowire with strong parabolic confinement in the $y$ and $z$ directions has the form~\cite{24}
\begin{equation}
    V_{C}^{1D}(x_1,x_2)=\frac{1}{\sqrt{2\pi}4\varepsilon_0\varepsilon l}\text{erfcx}\left(\frac{|x_1-x_2|}{\sqrt{2}l}\right).
\label{eq:Vc}
\end{equation}

The time evolution of the system is described by the time-dependent Schr\"odinger equation $i\hbar\frac{\partial}{\partial t}\Psi(x_1, x_2, t)={H}\Psi(x_1, x_2, t)$ which we solve numerically using an explicit Askar-Cakmak scheme~\cite{AC}. As initial condition we take the ground state $\Psi(x_1, x_2,t_0)=\Psi_0(x_1,x_2)$ for the two self-confined electrons under the metal electrodes which we obtain by solving the eigenvalue equation ${H}\Psi_0(x_1, x_2)=E\Psi_0(x_1, x_2)$ with the imaginary time propagation method~\cite{ITP}. In this  approach electron-electron correlations are taken into account exactly.
 
To characterize properties of the system we evaluate the spin density (i-th component) of the two-electron system,
\begin{eqnarray}
\nonumber
\rho_{S_i}(x,t)&=&\frac{\hbar}{2}\int_{x_L}^{x_R}dx'\big(\Psi^\dag(x,x',t)\sigma_i\otimes{I}\Psi(x,x',t)\\ 
&+&\Psi^\dag(x',x,t){I}\otimes\sigma_i\Psi(x',x,t)\big),
\label{eq:Srho}
\end{eqnarray}
where  $\sigma_i$ is i-th Pauli matrix, $i=x,y,z$. Consequently, with this formula the expectation value of the electron spin in the left ($s_z^{L}$) or right ($s_z^{R}$) part of the nanostructure takes the form,
\begin{equation}
s_z^{L(R)}(t)=\int_{x_L(x_0)}^{x_0(x_R)}dx\rho_{S_z}(x,t),
\label{eq:sLR}
\end{equation}
where $x_0$ is the midpoint of the nanowire and $x_L$ ($x_R$) is the left (right) end of the nanowire.
The amount of entanglement between the spin in the left and the right part of the nanowire can be quantified by calculating the  concurrence~\cite{PhysRevLett.78.5022, *PhysRevLett.80.2245} which is defined as $C(\rho^{S_{LR}})=\max\{0, \sqrt{\lambda_1}-\sqrt{\lambda_2}-\sqrt{\lambda_3}-\sqrt{\lambda_4}\}$.
It varies from zero for completely separable (non-entangled) states to unity for maximally entangled states. Here, $\lambda_i$ are eigenvalues (in decreasing order) of  $\tilde{\rho}=\rho_{S_{LR}}(\sigma_y\otimes\sigma_y){\rho_{S_{LR}}}^*(\sigma_y\otimes\sigma_y)$,
$\rho_{S_{LR}}$ is the reduced density matrix describing two-electron spin states in the left and right part of the nanowire.%

\begin{figure}[ht!]
\includegraphics[width=8.6cm]{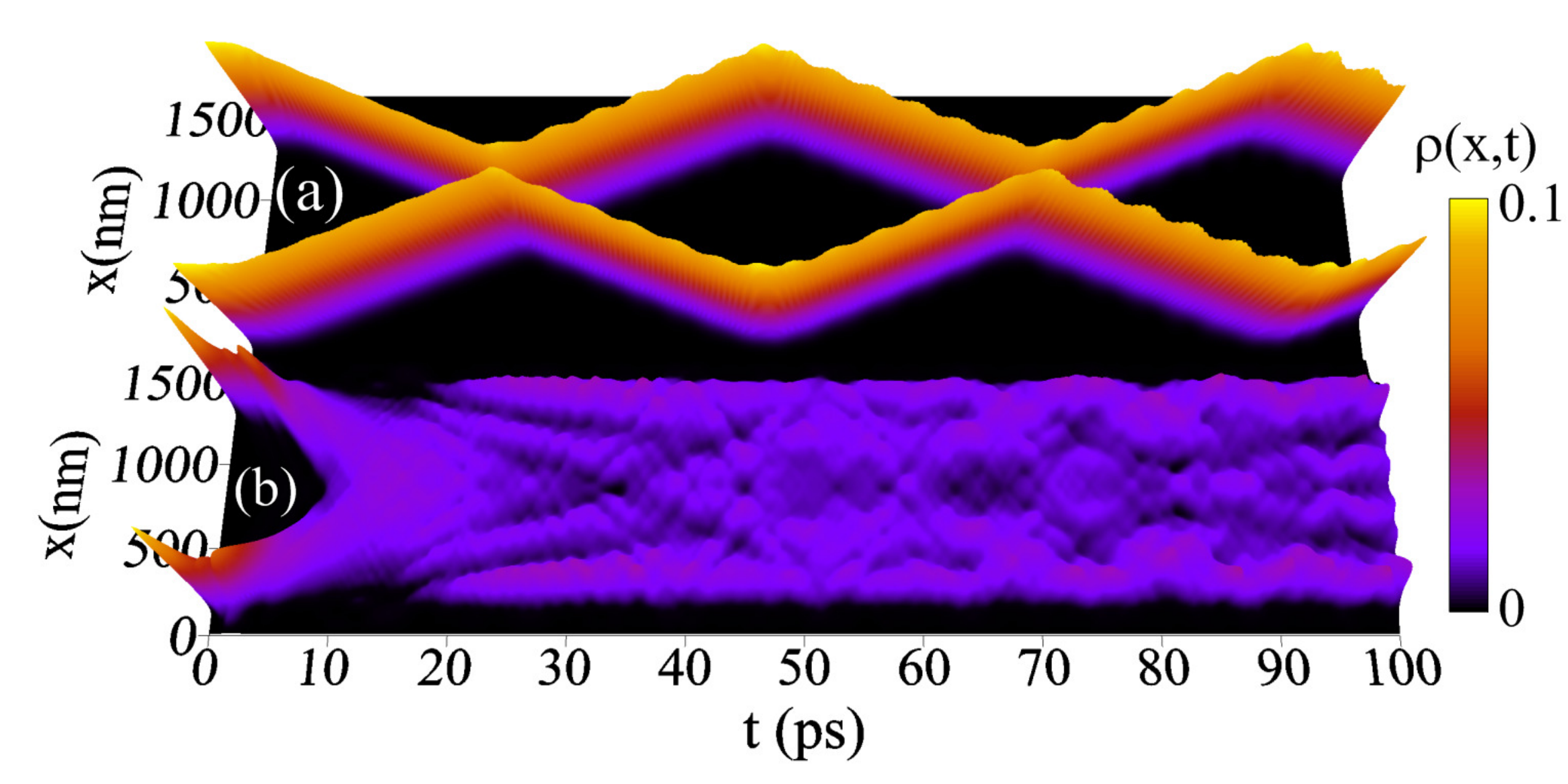}
\caption{\label{fig:coll2} Time evolution of the charge density $\rho(x,t)$ illustrating the difference in propagation and collision of two incident electrons between 
the case  with (a)  and without (b) image charge potential. Note the self-trapping and the soliton-like behavior in case (a).}
\end{figure}

{\it Results.} First we investigate the charge dynamics and illustrate differences in propagation between self-trapped soliton-like electrons and `freely' propagating electrons not interacting with the metal. 
Initially, electrons are confined in the nanowire   under the electrodes $e_L$ and $e_R$ (see Fig.~1), which is achieved by applying $V_{1,2,3,4}=-1$~meV and zero voltage to the other gates $V_{L,R,J}=0$.
The electrons are forced to move against each other by changing the voltage on electrodes $e_2$ and $e_3$ to $V_2=V_3=0$ and by lowering the voltage on electrodes $e_1$ and $e_4$ to $V_1=V_4=-1.5$~mV. 
The time evolution of the two-electron charge density $\rho(x,t)$ along the wire is depicted in Fig.~\ref{fig:coll2}. It can be seen that the electrons being self-trapped under the metal maintain their charge density shape while moving, which is a characteristic feature of soliton waves. Furthermore, the shape is not affected by the collision with other electrons [Fig.~\ref{fig:coll2}(a)]. 
However this is not the case for `free' electrons that are not interacting with the metal [Fig.~\ref{fig:coll2}(b)].



In order to realize a quantum $\sqrt{\text{SWAP}}$ gate using the nanodevice from Fig.~\ref{fig:nan}, we propose the following scheme. Initially (for $t_0$) each of the electrons is  localized under spatially separated electrodes $e_L$ and $e_R$, respectively. 
Since the electrons are significantly away from each other (in our numerical simulation by about 1.2$\mu$m) we assume that it is possible to prepare each of the electrons independently in a well-defined spin state, {\it i.e.}, the electron confined under the electrode $e_{L}$($e_{R}$) is in spin up $s_z^{L}(t_0)=\hbar/2$ (spin down $s_z^{R}(t_0)=-\hbar/2$) state. Thus in the representation $(2)$ the initial state has the following form~\cite{sep}: $\Psi_0(x,y)=(0,\varphi_{L}(x_1)\,\varphi_{R}(x_2),-\varphi_{R}(x_1)\,\varphi_{L}(x_2),0)^T$, where $\varphi_{R}(x_i)$ and  $\varphi_{L}(x_i)$ are the single electron ground state orbitals localized in the right (R) and in the left (L) dot. In this situation there is no entanglement between electron spins, {\it i.e.}, $C(\rho(t_0))=0$.

\begin{figure}[ht!]
\includegraphics[width=8.6cm]{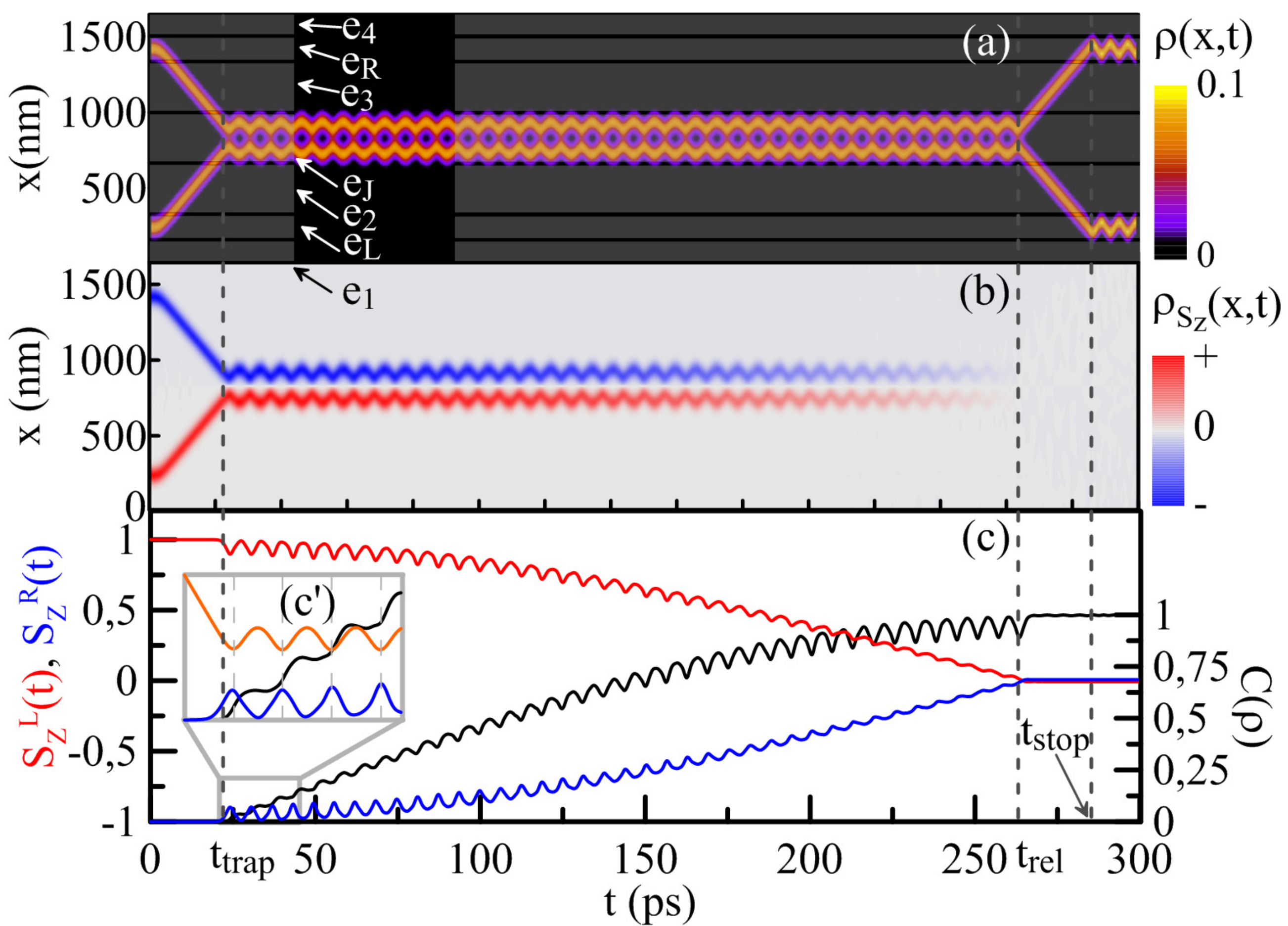}
\caption{\label{fig:sqrtswap} (a) Time evolution of the charge density $\rho(x,t)$ and scheme of the electrodes covering the nanodevice (grey).
 Time evolution of (b) the z-component of spin density $\rho_{S_z}$, (c) the z-component of expectation value of the electron spin in the left $s_L(t)$ (red) and the right $s_R(t)$ (blue) part of the nanowire (referred to the left axis). The concurrence $C$ (black) of the two-electron spin state
is shown with respect to the right axis in (c).
In the inset (c') of Fig.~3 (c) we plot  the average distance $|x_1-x_2|$ between electrons for the first few collisions.}
\end{figure}

Then, by changing the voltage on the electrodes (in the same manner as in the previous example), the electrons start to move. When the electrons reach the region under the electrode $e_J$ (for $t=t_{trap}\approx$~20ns) the voltage on the neighboring electrodes $e_2$ and $e_3$ is set to $V_{2}=V_{3}=$~-1mV, which traps the electrons under $e_J$. Then the electrons collide periodically under this electrode. 
The time evolution of the charge (z-th component of spin) density for the two electrons $\rho(x,t)$ ($\rho_{S_z}(x,t)$) is plotted in Fig.~\ref{fig:sqrtswap}(a) [Fig.~\ref{fig:sqrtswap}(b)]. The time evolution of the expectation value of the spin in the left and right part of the nanodevice and the concurrence $C$ is depicted in Fig.~\ref{fig:sqrtswap} (c). 
During each collision (due to exchange interaction which is intrinsically present in our model) electrons exchange a fraction of the spin and consequently entanglement builds up between electron spins in the left and right part of the nanodevice. This is illustrated in Fig.~\ref{fig:sqrtswap} (c') where we also plot the average distance between electrons (orange line). Dips in the average value of the spin or concurrence is due to a local and temporary increase of double occupation probability~\cite{PhysRevB.63.085311} during the soliton collision.

After many collisions for $t\approx 260ps$ we have a situation where the spin density vanishes and the spin in the left and the right dot is equal to zero, $s_z^{L}(t)\approx s_z^{R}(t)\approx 0$. Furthermore, the concurrence reaches $C(\rho(t))\approx 1$, which indicates that the spins are maximally entangled. However, the system is not yet in the spatially separated entangled state. In order to separate electrons 
from the area under the electrode $e_J$, the voltage on the electrodes $e_2$ and $e_3$ is switched (for $t_{rel}\approx$~260ps) to $V_2=V_3=0$, and after the last collision, the electrons start to move into opposite directions.
When they reach the region under the electrodes $e_L$ and $e_R$ (initial position) for $t_{stop}\approx 280$~ps, respectively, they are trapped again by changing the voltage on electrodes $e_{1,2,3,4}$ to $V_{1,2,3,4}=2$mV. Finally, the maximally spin-entangled state is obtained for spatially separated electrons characterized by the concurrence reaching $C(\rho)>0.999$.

\begin{figure}[ht!]
\includegraphics[width=8.6cm]{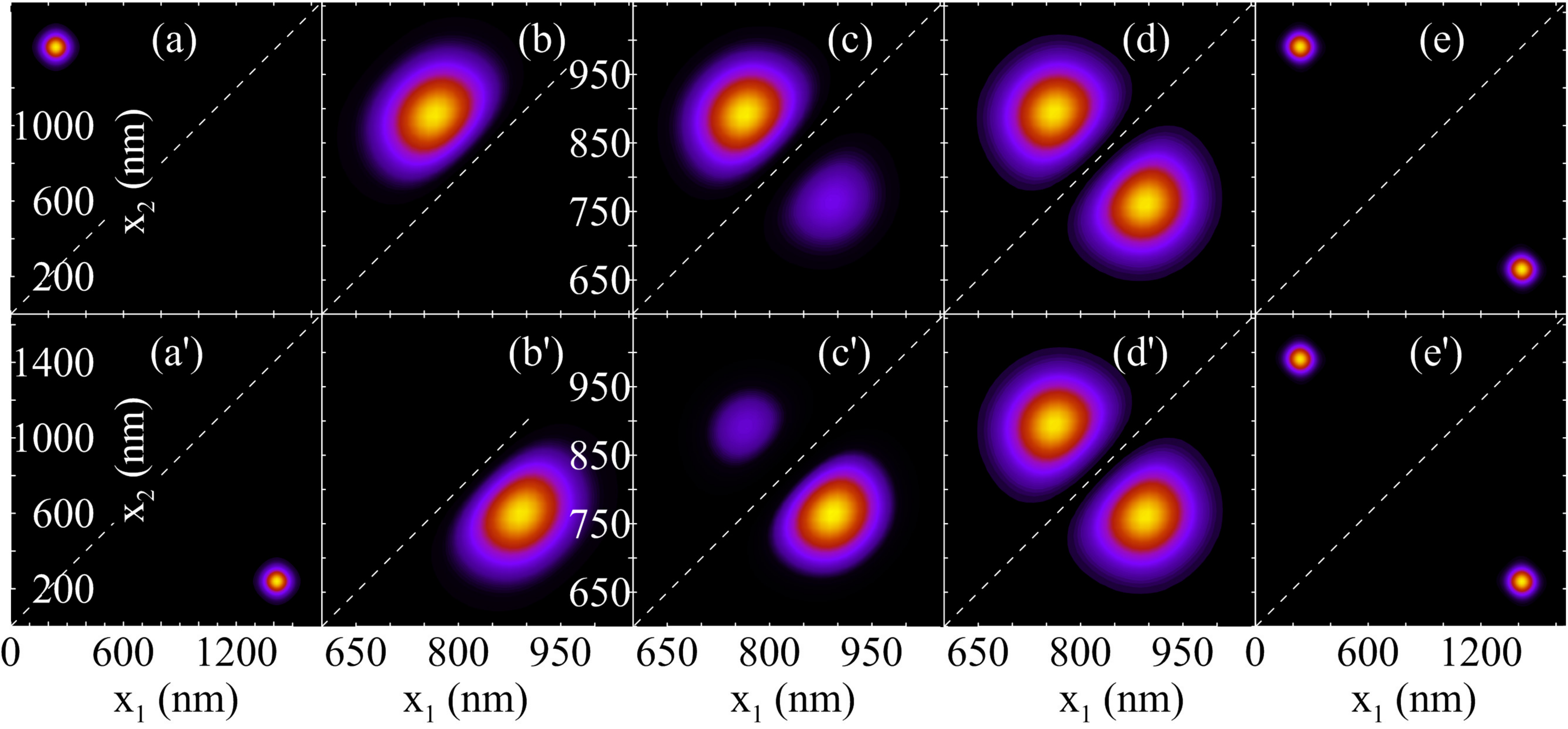}
\caption{\label{fig:cor} Probability densities  of the two-electron wave functions $|\psi_{\uparrow\downarrow}(x_1,x_2,t)|^2$ (upper panel)  and $|\psi_{\downarrow\uparrow}(x_1,x_2,t)|^2$ (lower panel with primes) for the instants (a) $t_0$, (b) $t_{trap}$, (c) $t_{1/2}$ [$C(\rho(t_{1/2}))=0.5$], (d) $t_{rel}$, and  (e) $t_{stop}$ during the realization of the $\sqrt{\text{SWAP}}$ gate.} 
\end{figure}

It is also instructive to analyze the probability densities
of the components of the total two-electron wave function for the case when the electrons spins are not entangled, partially entangled, and maximally entangled, respectively. For the chosen initial state with opposite spins, during the whole evolution components with parallel spins are zero $\psi_{\uparrow\uparrow}=\psi_{\downarrow\downarrow}=0$, while the other two are nonzero. The corresponding values  $|\psi_{\uparrow\downarrow}(x_1,x_2,t)|^2$ and $|\psi_{\downarrow\uparrow}(x_1,x_2,t)|^2$ are depicted in Fig.~\ref{fig:cor} for the initial moment $t_0$ - non-entangled spatially separated electrons, $t_{trap}$ - first collision, $t_{1/2}$ when the concurrence becomes $C(\rho(t_{1/2}))=0.5$, $t_{rel}$ last collision when the electron spins are maximally entangled, and finally for the maximally entangled and separated electron spins under electrodes $e_L$ and $e_R$.

\begin{figure}[ht!]
\includegraphics[width=8.6cm]{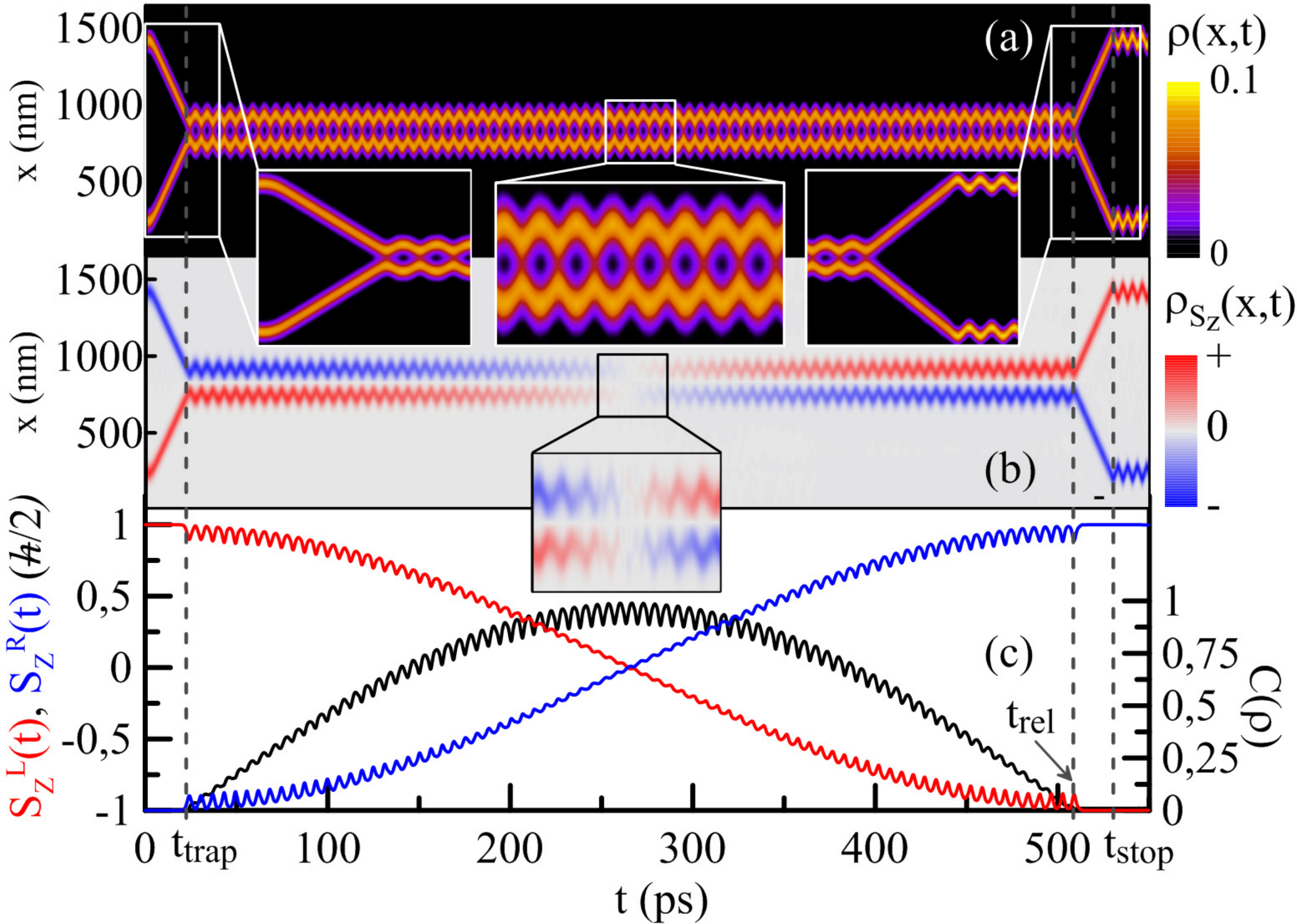}
\caption{\label{fig:swap}  Same as Fig.~\ref{fig:sqrtswap} but for the SWAP gate.}
\end{figure}

A similar procedure 
can be used to realize the two-qubit SWAP gate, which fully exchanges the spin of the two electrons. In this case electrons have to be released from under the electrode $e_J$ for $t_{rel}\approx 510$~ps and trapped again under $e_L$ and $e_R$ for $t_{stop}\approx 530$~ps. The results are shown in Fig.~\ref{fig:swap}. It is clearly seen how electrons exchange their spins during soliton collisions under electrode $e_J$. It is quite remarkable that despite many collisions the shape of the function (charge density) is still preserved and well localized.

The proposed scheme is most sensitive to proper adjustment of the $t_{trap}$ and $t_{stop}$ times. However, our method is quite robust 
against variations of $t_{rel}$, the releasing time of electrons from under $e_J$. For $t_{rel}-30\%t_{rel}<t<t_{rel}+30\%t_{rel}$ the concurrence reaches still more than $90\%$, $C(\rho(t))>0.9$. In our simulations the voltage switching time is about 4ps long. The operation times $\tau_{op}$ of the proposed gates are on the order of hundreds of picoseconds, which is three orders of magnitude shorter than the reported spin decoherence time in InAs QDs \cite{s1}. However, one can further tune (decrease) the gate operation time by increasing the voltage applied to the gates $e_3$ and $e_4$.

{\it Summary.} We have shown that the interplay of soliton-like properties of self-trapped electrons in gated semiconductor nanowires and the exchange interaction can be exploited to realize SWAP and $\sqrt{\text{SWAP}}$ gates.  The latter gate can be used to realize maximally entangled spin states of spatially separated electrons. The proposed gates act in an ultrafast manner (subnanoseconds) and are controlled only by small static voltages applied to the electrodes which makes our proposal particularly suitable for addressing individual spin qubits in scalable quantum registers.

We acknowledge support from the Swiss NSF, NCCR
QSIT, SCIEX (P.S.), and IARPA.
\nocite{*}
\bibliography{ent_sol} 

\end{document}